\documentclass[12pt]{article}
\usepackage{epsf}
\usepackage{url}
\usepackage{amsmath}
\usepackage{amssymb}
\usepackage{yfonts}

\begin{document}
%Status 10.6.2013
\begin{center}\large {\bf 
Do not circulate!} \end{center} 

\vspace{1cm} 
\Large \begin{center} {\bf Klein's ``Erlanger Programm'': do traces of
    it exist in physical theories?} \end{center} \normalsize  
 \begin{center} Hubert Goenner\\ Institute for Theoretical Physics\\
   University of G\"ottingen\\ Friedrich-Hund-Platz 1\\ D 37077
   G\"ottingen \end{center} 
\section{Introduction}

Felix Klein's ``Erlanger Programm'' of 1872 aimed at characterizing
geometries by the invariants of simple linear  transformation
groups.\footnote{We have noticed the difficult relationship of Sophus
  Lie with regard to F. Klein or W. Killing in connection with
  priority issues (\cite{Stub2002}, pp. 365-375). Klein acknowledged
  S. Lie as ``the godfather of my Erlanger Programm''
  (\cite{Klein1927}, p. 201). For the historical background of Lie
  groups, S. Lie and F. Klein cf. \cite{Rowe89}, \cite{Hawk1989}.} It
was reformulated by Klein in this way: ``Given a manifold[ness] and a
group of transformations of the same; to develop the theory of
invariants relating to that group''\footnote{``Es ist eine
  Mannigfaltigkeit und in derselben eine   Transformationsgruppe
  gegeben. Man soll  [..] die Theorie der   Beziehungen, welche {\it
    relativ} zur {\it Gruppe} invariant sind, untersuchen.'' - The
  translation given is by M. W. Haskell and authorized by Klein;
  cf. New York Math. Soc. 2, 215-249 (1892/93).} (\cite{Klein1893};
\cite{Klein1927}, p. 28). As if he had anticipated later discussions
about his program, a slightly different formulation immediately
preceding this is: ``Given a manifold[ness] and a group of
transformations of the same; to investigate the configurations
belonging to the manifoldness  with regard to such properties as are
not altered by the transformations of the group.''\footnote{``Es ist
  eine Mannigfaltigkeit und in derselben eine  Transformationsgruppe
  gegeben; man soll die der Mannigfaltigkeit angeh\"origen Gebilde
  hinsichtlich solcher Eigenschaften untersuchen, die durch die
  Transformationen der Gruppe nicht ge\"andert werden.'' (\cite{Klein1974}, pp. 34-35) - In a later annotation reproduced in \cite{Klein1921}, he denied as too narrow  an interpretation of his formulation strictly in the sense of looking only at algebraic invariants.} A wide interpretation of a later time by a mathematician is: ``According to F. Klein's viewpoint thus geometrical quantities like distance, angle, etc. are not the fundamental quantities of geometry, but the fundamental object of geometry is the transformation group as a symmetry group; from it, the geometrical quantities only follow'' (\cite{Schott95}, p. 39). On the other hand, by a physicist Klein's program is incorrectly given the expression ``[..] each geometry is associated with a group of transformations, and hence there are as many geometries as groups of transformations'' (\cite{Fersan84}, p. 2). The two quotations show a vagueness in the interpretation of F. Klein's ``Erlanger Programm'' by different readers. This may be due to the development of the concepts involved, i.e., ``transformation group'' and ``geometry'' during the past century. Klein himself had absorbed Lie's theory of transformation groups (Lie/Engel 1888-1893) when he finally published his Erlanger Program two decades after its formulation. Originally, he had had in mind linear transformations, not the infinitesimal transformations Lie considered.\\ 

F. Klein's point of view became acknowledged in theoretical physics at
the time special relativity was geometrized by H. Minkowski. Suddenly,
the Lorentz (Poincar\'e) group played the role Klein had intended for
such a group in a new geometry, i.e., in space-time. The invariants
became physical observables. But, as will be argued in the following,
this already seems to have been the culmination of a successful
application to physical theories of his program. What has had a lasting 
influence on physical theories, is the concept of {\em symmetry} as
expressed by (Lie-) transformation groups and the associated algebras with all their consequences. This holds particularly with regard to   conservation laws.\footnote{Important developments  following the
Noether theorems have been described by Y. Kosmann-Schwarzbach in her book about invariance and conservation laws \cite{Kosma2011}.} The reason is that in physical theories {\em fields} defined on the
geometry are dominant, not geometry itself. Also, for many physical
theories a geometry fundamental to them either does not exist or is
insignificant. A case in sight is the theory of the fractional quantum
Hall effect from which quasi-particles named ``anyons'' emerge. The
related group is the braid group describing topological transformations
\cite{HaKoWu1991}. What often prevails are geometrical models like the
real line for the temperature scale, or Hilbert space, an
infinite-dimensional linear vector space, housing the states of
quantum mechanical systems. In place of geometries, differential
geometrical ``structures'' are introduced. An example would be field 
re-parametrization for scalar fields in space-time. The fields can be 
interpreted as local coordinates on a smooth manifold. In the kinetic
term of the Lagrangian, a metric becomes visible which shows the
correct transformation law under diffeomorphisms. The direct
application of F. Klein's classification program seems possible only
in a few selected physical theories. The program could be replaced by
a scheme classifying the dynamics of physical systems with regard to
symmetry groups (algebras).      

The following discussion centers around finite-dimensional, continuous
groups. Infinite-dimensional groups will be barely
touched. (Cf. section \ref{section:conclusion}.) Also, the important 
application to discrete groups in solid state and atomic physics
(e.g., molecular vibration spectra) and, particularly, in
crystallography are not dealt with.\footnote{A survey of the groups
  is given in \cite{Klemm82}. For finite groups cf. also chapters 1 and
  2 of  \cite{KosSchwarz2010}. For the history of the interaction of
  mathematics  and crystallography cf. the book by E. Scholz
  \cite{Scholz1989}.} For the considerations to follow here, the question
need not be posed whether a reformulation of Klein's classifying idea
appropriate to modern mathematics is meaningful.\footnote{Some
  material in this respect may be found in P. Cartier's essay on the
  evolution of the concepts of space and symmetry \cite{Cartier2001}.}   

\section{Electrodynamics and Special Relativity}
It is interesting that F. Klein admitted that he had overlooked the
Galilei-group when writing up his ``Erlanger Programm'': ``Only the
emergence of the Lorentz group has led mathematicians to a more
correct appreciation of the Galilei-Newton group'' (\cite{Klein1927},
p. 56). It turned out later that the Galilean ``time plus space'' of
this group is more complicated than Minkowski's space-time
\cite{LevyLe63}, \cite{Kuenz1972}.   

What also had not been seen by F. Klein but, more than 30 years after
the pronounciation of the ``Erlanger Programm'', by mathematicians
E. Cunningham and H. Bateman, was that the Maxwell equations in vacuum
admit the 15-parameter conformal group as an invariance group
\cite{Bate1908}, (\cite{War2003}, p. 409-436, here p. 423). However,
this is a very specific case; if the electromagnetic field is coupled
to matter, this group is no longer admitted, in general.     

Special relativity, and with it Minkowski space, are thought to form a
framework for all physical theories not involving gravitation. Hence,
a branch of physics like relativistic quantum field theory in both its
classical and quantized versions is included in this application of
the ``Erlanger Programm''.\footnote{We recall that, on the strictest
mathematical level, an unambiguous union of quantum mechanics and
special relativity has not yet been achieved. Note also that algebraic
quantum field theory does not need full Minkowski space, but can get
along with the weaker light-cone structure supplemented by the
causality principle.} In the beginning of string theory (Veneziano
model), the string world sheet was likewise formulated in Minkowski
space or in a Lorentz space of higher dimension.   

We need not say much more concerning special relativity, but only
recall \linebreak Minkowski's enthusiasm about his new 
find: \begin{quote}``For the glory of mathematicians, to the infinite
  astonishment of remaining humanity, it would become obvious that
  mathematicians, purely in their fantasy, have created a vast area to
  which one day perfect real existence would be granted - without this
  ever having been intended by these indeed ideal chaps.'' (quoted
  from  (\cite{Klein1927}, p. 77).\footnote{``Es w\"urde zum Ruhme
    der Mathematik, zum grenzenlosen Erstaunen der \"ubrigen
    Menschheit offenbar werden, dass die Mathematiker rein in ihrer
    Phantasie ein gro{\ss}es Gebiet geschaffen haben, dem, ohne dass
    es je in der Absicht dieser so idealen Gesellen gelegen h\"atte,
    eines Tages die vollendete reale Existenz zukommen
    sollte.''}\end{quote} 
%In fact, any number of physicists now believe that space-time is
%really existing in the same sense as the room in which you are now,
%and is not just an elegant mathematical structure stored in our mind
%and helpful for describing nature.  
  
\section{General Relativity}
The description of the gravitational field by a Lorentz-metric, in
Einstein's general relativity, was predestinated to allow application
of Klein's program. The exact solutions of Einstein's field equations
obtained at first like the Schwarzschild- and de Sitter
solutions as well as the Einstein cosmos, defined geometries allowing
4- and 6-parameter Lie transformation groups as invariance
groups. Most of the exact solutions could be found just {\em because}
some invariance group had been assumed in the first place. Later, also
algebraical properties of the metrics were taken to alleviate the
solution of the non-linear differential equations. In the decades
since, it has become clear, that the {\em generic} solution of
Einstein's field equations does {\em not} allow an invariance group -
except for the diffeomorphisms Diff(M) of space-time M. As every
physical theory can be brought into a diffeomorphism-invariant form,
eventually with the help of new geometrical objects, the role of this
group is quite different from the one F. Klein had in mind.\footnote{Since
E. Kretschmann's papers of 1915 and 1917 \cite{Kret1915}, \cite{Kret1917}, there has been an extended discussion about an eventual physical content of the diffeomorphism group in general relativity; cf. \cite{Nort1992}, \cite{Nort1999}. It suffers from Einstein's identification of coordinate systems and physical reference systems with the latter being represented by tetrads (frames). These can be adapted  to matter variables.} He was well aware of the changed situation and saved his
program by reverting to {\em infinitesimal} point transformations. He
expressed his regret for having neglected, at the time of the
formulation of his ``Erlanger Programm'', Riemann's
Habilitations\-schrift of 1854 \cite{Riem1867}, and papers by
Christoffel and Lipschitz.\footnote{For the contributions of Lipschitz
  to the geometrization of analytical mechanics cf. (\cite{Luetz1999},
  pp. 29-31).} In fact, the same situation as encountered in general
relativity holds already in Riemannian geometry: generically, no
nontrivial Lie transformation group exists. Veblen had this in mind when
he remarked: \begin{quote} ``With the advent of  Relativity we became
  conscious that a space need not be looked at only as  a  `locus in
  which', but that it may have a structure, a field-theory of its
  own. This brought to attention precisely those  Riemannian
  geometries about which the Erlanger Programm said nothing, namely
  those whose group is the identity. [..]''
  (\cite{Veblen1928}, p. 181-182; quoted also by E. T. Bell \cite{Bell1945}, p. 443).\footnote{The original quote from Veblen continues with ``In such spaces there is essentially only one figure, namely the space structure as a whole. It became clear that in some respects the point of view of Riemann was more fundamental than that of Klein.'' } \end{quote} That general relativity allows only the identity as a Lie transformation group (in the sense of an isometry) to me is very much to the point. Perhaps, the situation is characterized best by H. Weyl's distinction between {\em geometrical   automorphisms} and {\em physical automorphisms} (\cite{Scholz2013}, p. 17). For general relativity, this amounts to Diff(M) on the one hand, and to the unit element on the other. Notwithstanding the useful identities following from E. Noether's second theorem, all erudite discussions about the physical meaning of Diff(M) seem to be
adornments for the fact that scalars are its most general invariants
possible on space-time. Usually, physical observables are transforming
covariantly; they need not be invariants. While the space-time metric
is both an intrinsically geometric quantity and a dynamical physical
field, it is not a representation of a finite-dimensional Lie
transformation group: F. Klein's program just does not apply. If
Einstein's endeavour at a unified field theory built on a more general
geometry had been successful, the geometrical quantities adjoined to
physical fields would not have been covariants with regard to a
transformation group in Klein's understanding.\\      

But F. Klein insisted on having strongly emphasized in his
program: ``that a point transformation $x_i = \phi(y_1...y_n)$ for an
infinitely small part of space always has the character of a linear
transformation [..]''\footnote{``da{ss} eine Punkttransformation [..]
  f\"ur eine unendlich kleine Partie des Raumes immer den Charakter
  einer linearen Transformation hat''.} (\cite{Klein1927}, p. 108). A 
symmetry in general relativity is defined as an isometry through
Killing's equations for the infinitesimal generators of a 
Lie-algebra. Thus in fact, F. Klein's original program is restricted
to apply to the tangent space of the Riemannian (Lorentz-)
manifold. This is how  E. Cartan saw it: a manifold as the envelope of
its tangent spaces; from this angle he developed his theory of groups
as subgroups of GL(n,R) with help of the concept of
G-structure.\footnote{H. Weyl with his concept of purely infinitesimal 
geometry in which a subgroup $G\subset SL(n,R)$ (generalized
"rotations") acts on every tangent space of the manifold,
separately, took a similar position (\cite{Scholz2012}, p. 24).}
Cartan's method for ``constructing finitely and globally inhomogeneous
spaces from infinitesimal homogeneous ones'' is yet considered by
E. Scholz as ``a reconciliation of the Erlangen program(me) and Riemann's
differential geometry on an even higher level than Weyl had
perceived''(\cite{Scholz2012}, p. 27).\footnote{For a Lie group $G \subset L$,
the homogeneous space corresponds to {\gothfamily l}/{\gothfamily
  g} $\cong T_pM$, where {\gothfamily l} and {\gothfamily g} are the
respective Lie algebras.}      

An extension of general relativity and its dynamics to a Lorentz-space
with one time and four space dimensions was achieved by the original
Kaluza-Klein theory. Its dimensional reduction to space-time led to
general relativity and Maxwell's theory refurbished by a scalar
field. Since then, this has been generalized  in higher dimensions to
a system consisting of Einstein's and the Yang-Mills equations
\cite{Kerner1968}, and also by including supersymmetry. An enlargement
of general relativity allowing for supersymmetry is formed by supergravity
theories. They contain a (hypothetical) graviton as bosonic particle with
highest spin 2 and its fermionic partner of spin 3/2, the
(hypothetical) gravitino; cf. also section \ref{section:susy}.  
 
\section{Phase space}    
A case F. Klein apparently left aside, is phase space parametrized by
generalized coordinates $q_i$ and generalized momenta $p_i$ of
particles. This space plays a fundamental role in statistical 
mechanics, not through its geometry and a possibly associated
transformation group, but because of the well known statistical
ensembles built on its decomposition into cells of volume $h^3$ for
each particle, with $h$ being Planck's constant. For the exchange of
indistinguishable particles with spin, an important role is played by
the permutation group: only totally symmetric or totally
anti-symmetric states are permitted. In 2-dimensional space, a
statistics ranging continuously between Bose-Einstein and Fermi-Dirac
is possible.  

The transformation group to consider would be the abelian group of
contact transformations (cf. \cite{Hoel1939}): \begin{equation}
  q^{\prime}_{i} = f_{i}(q_i, p_j)~,~p^{\prime}_j´= g_{j}(q_i,
  p_j)~,\end{equation} which however is of little importance in
statistical mechanics.\footnote{It is only loosely connected with Lie's geometric contact transformation which transforms plane surface elements into each other. Manifolds in contact, i.e., with a common (tangential) surface element remain in contact after the transformation. A class of linear differential equations is left invariant; cf. \cite{Klein1894}, pp. 19-20.} In some physics textbooks, no difference is made between contact and canonical transformations, cf. e.g.,   \cite{CorSte1957}. In others, the concept of phase space is limited to the cotangent bundle of a manifold with a canonical symplectic structure (\cite{AbraMar1978}, p. 341). An important subgroup of canonical
transformations is given by all those transformations which keep
Hamilton's equations invariant for {\em any} Hamiltonian.\footnote{For  contact transformations with higher derivatives cf. \cite{Yare1997}.} Note that for the derivation of the Liouville equation neither a Hamiltonian nor canonical transformations are needed. In this situation, symplectic geometry can serve as a model space with among others, the symplectic groups $SP(n, R)$ acting on it as transformation groups. Invariance of the symplectic form $ \Sigma_i^n (dq_i \wedge dp_i)$ implies the reduction of contact to canonical transformations. Symplectic space then might be viewed in the spirit of F. Klein's program. He does not say this but, in connection with the importance of canonical transformations to ``astronomy and mathematical physics'', he speaks of ``quasi-geometries in a $R_{2n}$ as they were developed by Boltzmann and Poincar\'e [..]'' (\cite{Klein1926}, p. 203).  

In analytical mechanics, Hamiltonian systems with conserved energy are
studied and thus time-translation invariance is
assumed. Unfortunately, in many systems, e.g., those named ``dynamical    
systems'', energy conservation does not hold. For them, attractors can
be interpreted as geometrical models for the ``local asymptotic
behavior'' of such a system while bifurcation forms a ``geometric
model for the controlled change of one system into another''
(\cite{AbraShaw1992}, p. XI). Attractors can display symmetries, e.g., 
discrete planar symmetries \cite{Cartal1998}, etc.

In statistical thermodynamics, there exist phase transitions between
thermodynamic phases of materials accompanied by ``symmetry
breaking''. As an example, take the (2nd order) transition from the
paramagnetic phase of a particle-lattice, where parallel and
anti-parallel spins compensate each other to the ferromagnetic phase
with parallel spins. In the paramagnetic state, the
full rotation group is a continuous symmetry. In the ferromagnetic
state below the Curie-temperature, due to the fixed orientation of the
magnetization, the rotational symmetry should be hidden: only axial
symmetry around the direction of magnetization should show up. However, in the 
Heisenberg model (spin 1/2) the dynamics of the system is rotationally
invariant also below the Curie point. The state of lowest energy
(ground state) is degenerate. The symmetry does not annihilate the
ground state. By picking a definite direction, the system
spontaneously breaks the symmetry with regard to the full rotation
group. When a continuous symmetry is spontaneously broken, massless
particles appear called Goldstone(-Nambu) bosons. They are corresponding 
to the remaining symmetry. Thus, while the dynamics of a system placed into a 
fixed external geometry can be invariant under a transformation group,
in the lowest energy state the symmetry may be reduced. This situation
seems far away from F. Klein's ideas about the classification of
geometries by groups.   
                      
\section{Gauge theories}
\label{section:YaMi}
Hermann Weyl's positive  thoughts about Klein's program were expressed
in a language colored by the political events in Germany at the
time: \begin{quote}``The dictatorial regime of the projective idea in  
  geometry was first broken by the German astronomer and geometer
  M\"obius, but the classical document of the democratic platform in
  geometry, establishing the group of transformations as the ruling
  principle in any kind of geometry, and yielding equal rights of
  independent consideration to each and every such group, is
  F. Klein's `Erlanger Programm'. ''(quoted from Birkhoff \& Bennet
  \cite{BirBen1988})\end{quote} 

Whether he remembered this program when doing a very important step
for physics is not known: H. Weyl opened the road to {\em gauge  
  theory}. He associated the electromagnetic 4-potential with a
connection, at first unsuccessfully by coupling the gravitational and
electrodynamic fields (local scale invariance). A decade later then,
by coupling the electromagnetic field to matter via Dirac's wave
function; for the latter he expressly invented 2-spinors. The
corresponding gauge groups were $R$ and $ U(1)$, respectively. This
development and the further path to Yang-Mills theory for 
non-abelian gauge groups has been discussed in detail by
L. O'Raifeartaigh and N. Straumann\footnote{The original Yang-Mills gauge theory
 corresponded to SU(2)-isospin symmetry of the strong interaction.}
\cite{ORai1997}, \cite{ORaiStrau2000}. Weyl had been convinced 
about an intimate connection of his gauge theory and general
relativity: ``Since gauge invariance involves an arbitrary function 
$\lambda$ it has the character of `general' relativity and can
naturally only be understood in that context'' (\cite{Weyl1929},
translation taken from \cite{ORaiStrau2000}). But he had not yet taken 
note of manifolds with a special mathematical structure introduced
since 1929, i.e., {\em fibre bundles}. Fibre bundles are local
products of a base manifold (e.g., space-time), and a group. The
action of the group creates a fibre (manifold) in each point of the
base. Parallel transport in base space corresponds to a connection defined in a
section of the bundle. In physics, the transformation group may be a
group of ``external'' symmetries like the Poincar\'e group or of
``internal'' symmmetries like a Yang-Mills (gauge) group. A well known
example is the frame bundle of a vector bundle with structure group
GL(n;R). It contains all ordered frames of the vector space (tangent
space) affixed to each point of the base manifold. Globally, base and
fibres may be twisted like the M\"obius band is in comparison with a
cylindrical strip.\footnote{Since the introduction of fibre spaces
  by H. Seifert in 1932, at least five definitions of fibre bundles
  were advanced by different researchers and research groups
  \cite{McCleary}. The first textbook was written by Steenrod
  \cite{Steen1951}.} In 1929, Weyl had not been able to see the gauge
potential as a connection in a principal fibre bundle. Until this was
recognized two to three decades had to pass.     

Comparing the geometry of principal fibre bundles with Riemannian
(Lorentzian) geometry, F. Klein's program would be realized in the
sense that a group has been built right into the definition of the
bundle. On the other hand, the program is limited because the group
can be any group. In order to distinguish bundles, different groups
have to be selected in order to built, e.g., $SU(2)-, SU(3)$-bundles,
etc.\footnote{Elementary particles are classified with regard to
  local gauge transformations $SU(3)_{c} \times SU(2)_{L} \times
  U(1)_{Y}$. The index $c$ refers to color-charge, $Y$ to weak
  hypercharge, and $L$ to weak isospin. For a review of the
  application of gauge theory to the standard model cf. \cite{Wein2005}.} This is a classification of bundle geometry in a similar sense
as isometries distinguish different Lorentz-geometries. To classify different
types of bundles is another story.\\ 

Moreover, in gauge theories, the relation between observables and
gauge invariants is not as strong as one might have wished it to
be. E.g., in gauge field theory for non-abelian gauge groups, the
gauge-field strength (internal curvature) does not commute with the
generators of the group: it is not an immediate observable. Only
gauge-invariant polynomials in the fields or, in the  quantized
theory, gauge-invariant operators are observables. In contrast, the
energy-momentum tensor is gauge-invariant also for non-abelian gauge groups.    

In terms of the symmetry\footnote{$SU(2) \times U(1)$ symmetry of
  electroweak interactions; approximate flavour $SU(3)$-symmetry of strong
interactions.}, gauge invariance is spontaneously broken, both in
the case of electroweak and strong interactions.

General relativity with its metric structure is not a typical gauge
theory: any external transformation group would not only act in the
fibre but also in the tangent space of space-time as well. Thus, an
additional structure is required: a soldering form gluing the tangent
spaces to the fibres \cite{Ehres}. Many gauge theories for the
gravitational field were constructed depending on the group chosen: 
translation-, Lorentz-, Poincar\'e, conformal group etc.\footnote{A
  recent reader about gauge theories of gravitation is
  \cite{BlagoHehl2013}.} We will come back to a Poincar\'e gauge
theory falling outside of this Lie-group approach in section
\ref{section:furdev}.\\    
   
\section{Supersymmetry}
\label{section:susy}
Another area in physics which could be investigated as a possible
application of Klein's program is supersymmetry-transformations and 
supermanifolds. Supersymmetry expressed by super-Lie-groups is a
symmetry relating the Hilbert spaces of particles (objects) obeying
Bose- or Fermi-statistics (with integer or half-integer spin-values, 
respectively). In quantum mechanics, anti-commuting supersymmetry operators exist mapping the two Hilbert spaces into each other. They commute with the Hamiltonian. If the vacuum state (state of minimal energy) is annihilated by the supersymmetry operators, the
1-particle states form a representation of supersymmetry and the total
Hilbert space contains bosons and fermions of {\em equal}
mass.\footnote{For the geometry of supersymmetric quantum mechanics
  cf. e.g., \cite{Witten1982}. There, supersymmetric quantum field
  theory is formulated on certain infinite-dimensional Riemannian 
  manifolds.} As this is in contradiction with what has been found,
empirically, supersymmetry must be broken (spontaneously) in nature.   

For an exact supersymmetry, the corresponding geometry would be
supermanifolds, defined as manifolds over superpoints, i.e., points
with both commuting coordinates as in a manifold with n space
dimensions, and anti-commuting ``coordinates'' forming a
Grassmann-algebra $\zeta, \bar{\zeta}$ (``even'' and ``odd'' elements)
\cite{deWitt1984}.\footnote{As a supermanifold is not only formed from
  the usual points with commuting coordinates, another definition has
  been used: It is a topological space with a sheaf of superalgebras
  ($Z_{2}$-graded commutative algebras).} As a generalization of
Minkowski space, the coset space Poincar\'e/Lorentz in which the
super-Poincar\'e group acts, is called superspace (\cite{Freund1986},
Chapter 6), \cite{Sohn1985}, p. 107). Superspace is a space with 8
``coordinates'' $z^{A}=(x^{k},
\theta^{\mu},\bar{\theta}_{\dot{\mu}})$, where $x^{k}$ are the usual 
real space-time coordinates plus 4 real (anti-commuting) ``fermionic'' 
coordinates from a Weyl-spinor $ \theta^{\mu}$ and its conjugate
$\bar{\theta}_{\dot{\mu}})$.  

A super-Lie group $G$ is a Lie group with two further properties: 1)
it is a supermanifold the points of which are the group elements of
$G$ ; 2) the  multiplicative map F: $G \rightarrow G \times G $ is
differentiable (\cite{deWitt1984}, p. 123).\footnote{According to
  (\cite{deWitt1984}, p. 173-174) {\em conventional} super-Lie groups
  and {\em unconventional} super-Lie groups unrelated to graded
  algebras must be distinguished.} All classical Lie groups have
extensions to super-Lie groups. Most important for quantum field
theory is the super-Poincar\'e group and its various associated
super-Lie-algebras. The super-Poincar\'e algebras contain both
Lie-brackets and anti-commuting (Poisson) brackets. A superparticle
(supermultiplet) corresponds to a reducible representation of the
Poincar\'e algebra. 

The {\em geometry of supermanifolds} seems to play only a minor role
in physics. An example for its use would be what has been called the
gauging of supergroups \cite{Nieuw85}. Local super-Lie algebras are
important because their representations constitute superfields by
which the dynamics of globally or locally supersymmetric physical
theories like {\em supergravity} are built.\footnote{Superfields can
  be defined as   functions on superspace developed into power series
  in the nilpotent   Grassmann-variables in superspace; the power
  series break off after   the term  $\theta \theta \bar{\theta}
  \bar{\theta}~ a(x)$. Local   supersymmetric theories are theories
  invariant under {\em supergauge}-transformations.} Supergravity
containing no particle of spin larger than 2 can be formulated in
Lorentz-spaces up to maximal dimension 11. In space-time, at least 7
supergravities can be formulated. Yet, a geometrical construct like a
supermetric is of no physical importance.\\ 
 
This all too brief description is intended to convey the idea that, in
physics, the role of supersymmetry primarily is not that of a
transformation group in a supermanifold but of a group restricting the
dynamics of interacting fields. By calling for invariants with regard to
supersymmetry, the choice of the dynamics (interaction terms in the
Lagrangian) is narrowed considerably. The supersymmetric
diffeomorphism group can be used to formulate supersymmetric theories
in terms of differential forms on superspace: ``superforms''
(\cite{WesBag1983}, Chapter XII). Possibly, B. Julia envisioned the
many occuring supersymmetry groups when drawing his illustration
for supergravities ``A theoretical cathedral'' and attaching to the
x-axis the maxim: GEOMETRY $\simeq$ GROUP THEORY (\cite{Julia1985}, p. 357). When the view is narrowed to F. Klein's ``Erlanger Programm'' as is done here, then the conclusion still is that the program cannot fare better in supergravity than in general relativity.    

\section{Enlarged Lie algebras}
\label{section:furdev}
We now come back to space-time and to a generalization of
(Lie)-transformation groups acting on it. As insinuated before, for
the classification of structures in physical theories the attention
should lie rather on the algebras associated with the groups;
geometrical considerations intimately related to groups are of little
concern. Lie algebras have been generalized in a number of ways. One
new concept is ``soft'', ``open'' or ``nonlinear'' Lie algebras, in
which the structure constants are replaced by structure functions
depending on the generators themselves. They can also be interpreted as
infinite-dimensional Lie algebras (\cite{FuSchwei1997}, pp. 60-61). An
example from physics are local supersymmetry transformations
(defined to include diffeomorphisms, local Lorentz and local
supersymmetry transformations) which form an algebra with structure
{\em functions}. They depend on the symmetry generators themselves
(\cite{Nieuw85}, p. 140).\\   
Another generalization is ``local Lie algebras'' which arise as the
Lie algebras of certain infinite-dimensional Lie groups. The structure
of the Lie algebra in given by: \begin{equation}  [f_1, f_2] = 
  \Sigma_{i,j,k}^{n} c^{ij}_{~k}x^{k} \partial_{i} f_1  \partial_{j}
  f_2~, \nonumber \end{equation} where $f_1, f_2$ are smooth functions
on a smooth manifold, $ \partial_{k}$ the partial derivatives with
respect to local coordinates on $M$, and $c^{ij}_{~k}$ the structure 
constants of an n-dimensional Lie algebra (cf. \cite{Berez1967}, section
7). This seems to be a rather special kind of algebra. 

Recently, a further enlargment has been suggested called ``extended
Lie algebras'' and in which the structure constants are replaced by
functions of the space-time coordinates. In the associated groups, the
former Lie group parameters are substituted by arbitrary functions
\cite{Goen2013}. The Lie algebra elements form an ``involutive
distribution'', a smooth distribution $V$ on a smooth manifold
$M$. The Lie brackets constitute the composition law; the injection $V
\hookrightarrow TM$ functions as the anchor map. Thus, this is a
simple example for a {\em tangent Lie algebroid}. In addition to the
examples from physics given in \cite{Goen2013}, the Poincar\'e gauge
theory of F.-W. Hehl et al. seems to correspond to the definition of
an  extended Lie  algebra. In this theory, the difference with the Lie
algebra of the Poincar\'e group is that the structure functions now
contain the frame-metric and the gauge fields, i.e., curvature as
rotational and torsion as translational  gauge field, all dependent on
the space-time coordinates \cite{HeHeKer1976}. 

\section{Conclusions}
\label{section:conclusion}
In the course of ranging among physical theories with an eye on
F. Klein's ``Erlanger Programm'', we noticed that the focus had to be
redirected from groups and geometry to algebras and the dynamics of
fields. In particular, with regard to infinite-dimensional groups, the
discussion within physical theories of Klein's program would have been easier
had it been formulated in terms of algebras. Then, also Virasoro- and
Kac-Moody algebras, appearing among others in conformal (quantum)
field theory and in string theory could have been included in the
discussion.\footnote{The Virasoro group is an infinite dimensional
  group related to conformal (quantum) field theory in 2
  dimensions. It is defined as $Diff(S^1)$ where $S^1$ is the unit
  circle, and its geometry the infinite dimensional complex manifold
  $Diff(S^1)/S^1$ \cite{Zumin1988}. In string theory, the
  Virasoro-{\em algebra} appears. It is a central extension of a
  Witt-algebra providing unitary representations. The Witt-algebra is
  the Lie-algebra of smooth vector fields on $S^1$.} Hopf-algebras
occuring in {\em non-commutative geometry} could have formed another
example. With the mentioned change in focus included, the application
of Klein's program to physical theories is far more specific than a 
loosely defined methodological doctrine like the ``geometrization of
physics'' (cf. \cite{Kalin1988}). While both, general relativity and
gauge theory, can be considered as geometrized, they only partially
answer F. Klein's ``Erlanger Programm''. In physical theories, the
momentousness of Lie's theory of transformation groups easily
surpasses Klein's classification scheme.  
\section{Acknowledgments}
For the invitation to contribute to this volume and for his helpful
comments I thank A. Papadopoulos, Strasbourg. Inspiring discussions
with my colleague Detlef Buchholz, G\"ottingen, are gratefully
acknowledged. My sincere thanks go to Profs. Friedrich Hehl, K\"oln
and Erhard Scholz, Wuppertal whose comments contributed much to the
improvement of a first version.

\end{document}